\documentclass[aps,twocolumn,pra,superscriptaddress,eps2pdf]{revtex4-1}

\usepackage{epsfig}
\usepackage{amsfonts}
\usepackage{amssymb}
\usepackage{mathrsfs}
\usepackage{theorem}
\usepackage{amsmath}
\usepackage{times}
\usepackage{color}
\usepackage{soul}
\usepackage{rotating}
\usepackage{pbox}
\usepackage{array,float}
\usepackage{longtable}
\usepackage{lipsum}

\newcolumntype{L}{>{\centering\arraybackslash}m{3cm}}

\usepackage{ifpdf}
\ifpdf
\usepackage{epstopdf}   
\fi

\newtheorem{theorem}{Theorem}

\newtheorem{lem}{Lemma}

\theoremstyle{definition}

\def\id{\mbox{\small 1} \!\! \mbox{1}}

\def\id{\mbox{\small 1} \!\! \mbox{1}}

\def\id{{\mathchoice {\rm 1\mskip-4mu l} {\rm 1\mskip-4mu l} {\rm 1\mskip-4.5mu l} {\rm 1\mskip-5mu l}}}

\begin{document}

\title{Shorter gate sequences for quantum computing by mixing unitaries}
\author{Earl Campbell}
\email{earltcampbell@gmail.com}
\affiliation{Department of Physics and Astronomy, University of Sheffield,
Sheffield, UK
}

\begin{abstract}
Fault-tolerant quantum computers compose elements of a discrete gate set in order to approximate a target unitary.  The problem of minimising the number of gates is known as gate-synthesis.  The approximation error is a form of coherent noise, which can be significantly more damaging than comparable incoherent noise.  We show how mixing over different gate sequences can convert this coherent noise into an incoherent form.  As measured by diamond distance, the post-mixing noise is quadratically smaller than before mixing, without increasing resource cost upper bounds.  Equivalently, we can look for shorter gate sequences that achieve the same precision as unitary gate-synthesis. For a broad class of problems this gives a factor $1/2$ reduction in worst-case resource-costs.
\end{abstract}

\maketitle

The constraints of fault-tolerant quantum computing mean that the available quantum gates form a discrete set.  Such a gate set is said to be universal if it generates a group that gives a dense cover over all unitaries.  That is, any target unitary can be approximated to any desired level of precision with a sufficiently long sequence of gates.  The Solovay-Kitaev~\cite{kitaev02,dawson05,Fowler11,pham13} theorem ensures that whenever we have a universal gate set, we can achieve a circuit depth that is poly-logarithmic in the inverse precision.  The Solovay-Kitaev theorem is a very powerful and general result, but in practice yields very long gate sequences.  Remarkable progress beyond Solovay-Kitaev has been made in recent years by focusing on gate-sets that naturally arise in fault-tolerant quantum computing, in particular the Clifford+$T$ gate set, with the flourishing topic becoming known as gate-synthesis~\cite{kliuchnikov13,amy13,RS14,bocharov15}.

A common feature of both new and old approaches to gate-synthesis is the approximation of the target unitary with a different unitary.  Then the approximation error is a form of coherent noise, which has attracted attention as being especially pernicious to quantum computations~\cite{sanders15,kueng15}.   It has, however, been observed several times that mixing over equivalent circuits can average out coherent noise into less damaging incoherent noise~\cite{knill04,kern05,kliuchnikov12,wallman15,Knee15}.  For instance, when the individual gates suffer from coherent noise,  randomized compiling has been shown to quadratically reduce this noise source~\cite{wallman15}.  In the context of gate synthesis, the approximation error appears even when the components of our gate set are perfect, and so a different approach is required.  

Here we give the first general set of tools for mixing out the approximation errors in gate synthesis.  Quantifying this noise by the diamond norm, we find our approach reduces noise from $\epsilon$ to $O(\epsilon^2)$, without increasing the any worst-case metric of resource cost.  To be clear, by worst-case resource cost we mean the tightest available upper bound on resource cost. Alternatively, we can achieve $O(\epsilon)$ noise with reduced worst-case resource cost.  If the worst-case resource cost of unitary gate-synthesis scales as $A \mathrm{log}(\epsilon^{-1})^{\gamma} $, then using quantum channels $\epsilon$ noise can be attained with resource costs upper bounded by $A (1/2)^{\gamma}  \mathrm{log}(\epsilon^{-1})^{\gamma}$ in the small $\epsilon$ limit.  Many recent gate-synthesis algorithms have $\gamma=1$ scaling, and so in these setting we cut worst-case costs in half.  This is an extension of the notion of magic state dilution in Ref.~\cite{campbell16}, but here applied to synthesis of operations, rather than states.   When completing this work, some similar insights were reported by Hastings~\cite{Hastings15}, though without the explicit convex hull finding algorithm provided here.

\section{Notation}

We use $||\ldots||$ throughout for the operator norm, so that $||X||$ is the largest singular value of $X$.  We also make use of the Schatten 1 norm on operators denoted $|| \ldots ||_{1}$, which equals the sum of the singular values.  Throughout we make use of several norm properties discussed in standard texts~\cite{HJ01b,bhatia13}.  For a quantum channel we use the diamond norm $||\ldots||_{\diamond}$ where
\begin{equation}
	|| \mathcal{E} ||_{\diamond} := \mathrm{sup} \{ || (\mathcal{E}\otimes \id)(X) ||   ;  ||X||_1 \leq 1 \}.
\end{equation}
The diamond norm induces the diamond distance between two channels $\mathcal{E}$ and $\mathcal{E}'$, so that 
\begin{equation}
	d_{\diamond}(\mathcal{E}, \mathcal{E}'):=	\frac{1}{2}|| \mathcal{E} -  \mathcal{E}' ||_{\diamond},
\end{equation}
and is widely used~\cite{kitaev97} to quantify how well an imperfect channel $\mathcal{E}'$ approximates an ideal, target channel $\mathcal{E}$.  The diamond distance is well behaved under composition of channels, allowing it to be used in rigorous proofs, including proofs of the threshold theorem for fault-tolerant quantum computing~\cite{aharonov97}.  Despite the average fidelity gaining popularity and being easily measurable by randomised benchmarking~\cite{emerson05,emerson07,Knill08,Dankert09}, various commentators have observed that average fidelity is less meaningful than the diamond distance~\cite{sanders15}. 

In inexact gate synthesis, a sequence of available gates are composed to produce some $U$ that gives a good approximation to a target unitary $V$.  Techniques for gate synthesis typically report the precision of these approximations by taking $U-V$ and evaluating some norm.  This prompts us to ask how this notion of precision corresponds to the more versatile diamond distance.  Denoting, $\mathcal{U}$ and $\mathcal{V}$ as the channels corresponding to $U$ and $V$, we have
\begin{equation}
    	 d_\diamond( \mathcal{U} , \mathcal{V} ) \leq 	 || U - V || ,
\end{equation}
as shown in Refs.~\cite{Wang13,Wang15}.  In general, there is no simple lower bound. For instance, if $U=-V$ then $|| U - V ||=2$, but $\mathcal{U}=\mathcal{V}$ and so  $d_\diamond( \mathcal{U} , \mathcal{V} ) =0$.  However, these pathologies only arise when $|| U - V ||$ is large, and many families of  unitaries are well behaved.  Consider, for instance,  unitaries of the form $U=e^{i \theta Z}$ and $V=e^{i \theta' Z}$, for small $|\theta-\theta'|$ we find  $|| U- V ||$ is very close to the diamond distance~(see App. B of Ref.~\cite{campbell16} for more a more detailed discussion).  So while unitary precision and diamond distance are very different measures, they often coincide.  

Throughout we will use $\mathcal{G}$ to denote the available gate set, and $\mathfrak{C}: \mathcal{G} \rightarrow \mathbb{R}^+$ for the associated cost function.  To assess the depth of a circuit we would use a constant cost function $\mathfrak{C}(V)=1$ for all $V \in \mathcal{G}$.  However, for the Clifford+$T$ gate set the $T$ gates can be significantly more expensive than Clifford gates due to the resource overhead of magic state distillation~\cite{BraKit05,Bravyi12,Meier13,Jones13}.  In this setting, one often takes $\mathfrak{C}(T)=1$ and $\mathfrak{C}(C)=0$ for all $C$ in the Clifford group. The cost of a gate sequence is then taken to be the numerical sum of the composite gate costs.   We also use $\langle \mathcal{G} \rangle$ for the group generated by set $\mathcal{G}$.   We say a gate set is finite when $\mathcal{G}$ contains a finite number of elements.   Lastly, we will use $\mathrm{Conv}[\ldots]$ to denote the convex hull of a set of operators.

\section{Results}

Here we present two main results of this paper
\begin{theorem}
\label{genThm} 
Let $\mathcal{L}$ be some $d$ dimensional Lie group, which is a subgroup of a unitary group $SU(D)$. Let $\mathcal{G}$ be a finite gate set with cost function $\mathfrak{C}: \mathcal{G} \rightarrow \mathbb{R}^+$, such that $\langle \mathcal{G} \rangle$ is a dense cover of $\mathcal{L}$ and $\langle \mathcal{G} \rangle \subset \mathcal{L}$.  Assume we have a unitary synthesis algorithm: for every $V \in \mathcal{L}$ and all $\epsilon > 0$ the algorithm outputs a finite sequence $U = W_1 W_2 \ldots W_N \in \langle \mathcal{G} \rangle$, such that
\begin{align}
	|| U - V || & \leq \epsilon , \\
	\sum_{j=1}^N \mathfrak{C}(W_j) & \leq f(\epsilon) ,
\end{align}
where $f$ is the worst case cost of the unitary synthesis algorithm. It follows that we can construct a channel of the form 
\begin{equation}
		\mathcal{E}(\rho) =\sum_{j=1}^{n} p_j U_j \rho U^\dagger_j ,
\end{equation}
where all $U_j  \in \langle \mathcal{G} \rangle $ and each have cost upper bounded by $f(\epsilon)$, and provided $\epsilon <  0.01$ the post-mixing noise satisfies
\begin{equation}
\label{GenDD}
		d_\diamond( \mathcal{E} , \mathcal{V} ) \leq 10 \epsilon^2  .
\end{equation}
Therefore, $O(\epsilon^2)$ error in the diamond norm.
\end{theorem}
The simplest setting is that $\mathcal{L} = SU(D) $, so $d=D$, but we also allow for subgroups with $d<D$.  Few gate-synthesis techniques exist for multi-qubit or qudit problems, but our results apply there also. It directly applies to the familiar problem of performing general single-qubit rotations from the Clifford+$T$ gate set.  The natural cost function of this gate set is $\mathfrak{C}(T)=1$ and $\mathfrak{C}(C)=0$ for all $C$ in the Clifford group.  For such a cost function, Ross and Selinger~\cite{RS14} showed that efficient gate synthesis of any single qubit gate is possible with $f_{\mathrm{RS}}(\epsilon) = 9 \log_2 (\epsilon^{-1}) + O(\log_2 ( \log_2 (\epsilon))) $.  Using quantum channels, and no more gates, we can ensure $10 \epsilon^2$ precision in diamond distance.  

We use the terminology axial rotation for single qubit rotations about the $Z$ axis, and denote the group $\mathcal{L}_{\mathrm{ax}}$.  For such rotations the above findings apply with the function $f_{\mathrm{RS}}$.  However, the Ross and Selinger algorithm can generate axial rotations at a slightly lower cost with leading order $3 \log_2 (\epsilon^{-1})$, and other algorithms have been tailored to this special case.  So one might anticipate that resource savings could be made by tailoring our approach to axial rotations.  We find this is indeed the case, but we cannot blindly apply the above result to algorithms for axial rotations.  Note that Thm.~\ref{genThm} does not apply in this setting since the generated group $\langle \mathcal{G} \rangle$ contains gates outside $\mathcal{L}_{\mathrm{ax}}$.  That is, with $\mathcal{G}$ as the Clifford+$T$ set, the generated group has gates outside the axial rotation group, so $\langle \mathcal{G} \rangle \not \subset \mathcal{L}_{\mathrm{ax}}$.   However, our techniques are straightforwardly extended to such scenarios. 

\begin{theorem}
\label{axialThm} 
Let $\mathcal{L}_{\mathrm{ax}}$ be the group of axial rotations. Let $\mathcal{G}$ be a gate set with cost function $\mathfrak{C}: \mathcal{G} \rightarrow \mathbb{R}^+$ with Pauli $Z \in \mathcal{G}$ and $\mathfrak{C}(Z)$=0.  Assume we have a unitary synthesis algorithm: for every $V \in \mathcal{L}_{\mathrm{ax}}$ and all $\epsilon > 0$ the algorithm outputs a finite sequence $U = W_1 W_2 \ldots W_n \in \langle \mathcal{G} \rangle$, such that
\begin{align}
	|| U - V || & \leq \epsilon , \\
	\sum_{j=1}^N \mathfrak{C}(W_j) & \leq f_{\mathrm{ax}} (\epsilon) ,
\end{align}
where $f_{\mathrm{ax}}$ is the worst case cost of the unitary synthesis algorithm. It follows that we can construct a channel of the form 
\begin{equation}
		\mathcal{E}(\rho) =\sum_{j=1}^{4} p_j U_j \rho U^\dagger_j ,
\end{equation}
where all $U_j  \in \langle  \mathcal{G} \rangle $ and each have cost upper bounded by $f_{\mathrm{ax}}(\epsilon)$, and provided $\epsilon <  0.01$ the post-mixing noise satisfies
\begin{equation}
\label{AxialDD}
		d_\diamond( \mathcal{E} , \mathcal{V} ) \leq 5 \epsilon^2 .
\end{equation}
Therefore, $O(\epsilon^2)$ error in the diamond norm.
\end{theorem}

This result has a slightly better  $5 \epsilon^2$ instead of $10 \epsilon^2$, but more importantly benefits from using $f_{\mathrm{ax}}$ which gives a smaller resource overhead than for general qubit rotations.

Let us reflect on how this free error suppression can be swapped in exchanged for cheaper gate sequences.  We instead run  our protocol and use gate sequences of cost not exceeding  $f(\sqrt{\epsilon / \alpha })$, where $\alpha$ is $5$ or $10$ depending on which theorem we employ.  It follows that the post-mixing noise is bounded by $\epsilon$, but worst-case resource costs are reduced.  However, in a particular instance of a problem the resource cost could be much less than the worst-case cost.  As such, whenever a new protocol offers a superior worst-case cost, there is no ironclad promise that the protocol will have a lower resource cost in all problem instances, though such anomalies are probably quite rare.  We proceed on the mild assumption that improved worst-case resource costs accurately reflect actual resource savings, and next give a precise account of this saving.

\begin{figure}
	\includegraphics{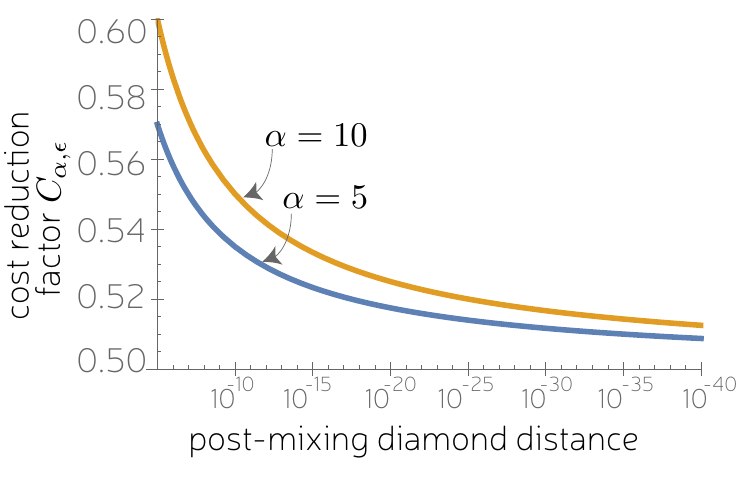}
\caption{The resource savings of our approach over unitary gate-synthesis is $C_{\alpha, \epsilon}^\gamma$ and here we show $C_{\alpha, \epsilon}$ (see Eq.~\eqref{Cval}) for $\alpha=5,10$ and a range of post-mixing error rates.  The different $\alpha$ correspond to different constant factors in Eq.~\eqref{GenDD} and Eq.~\eqref{AxialDD}. }	
\label{fig::conv}
\end{figure}

The form of $f$ for unitary gate-synthesis is typically $f(\epsilon) \sim A \log (\epsilon^{-1})^\gamma$ upto a small $O\{ \log [\log (\epsilon^{-1})  \}$ contribution.  Our reduced cost is then
\begin{align}
\label{EQ::resource}
		f(\sqrt{\epsilon / \alpha }) & \sim A \log ( (\epsilon / \alpha )^{-1/2} )^\gamma  \\ \nonumber
	& \sim  A \left( \frac{ \log (\epsilon^{-1})+  \log (\alpha)}{2} \right)^\gamma  \\ \nonumber
	& \sim  A \log (\epsilon^{-1})^\gamma \left[\left(\frac{1}{2}\right)\left( 1 + \frac{\log (\alpha)}{\log (\epsilon^{-1})} \right)\right]^\gamma  .
\end{align}
Therefore, our resource savings are a factor $C_{\alpha, \epsilon}^{\gamma}$ where 
\begin{align}
\label{Cval}
	C_{\alpha, \epsilon} & = 	\left(\frac{1}{2}\right)\left( 1 + \frac{\log (\alpha)}{\log (\epsilon^{-1})} \right)
\end{align}
collects the terms in the square bracket of Eq.~\eqref{EQ::resource}.  In the small $\epsilon$ limit we have $C_{\alpha, \epsilon} \rightarrow 1/2$.  Typically, $\epsilon$ is very small with many algorithms requiring $\epsilon\ll 10^{-10}$ and so $C_{\alpha, \epsilon} \sim 1/2$ is a reasonable approximation.  Convergence toward $1/2$ is shown in Fig.~\ref{fig::conv}, with the speed of convergence dictated by $\alpha$.  When proving our theorems we focus on clarity rather than minimising $\alpha$ and believe smaller $\alpha$ is plausible.  Lastly, recall that for single qubit problems known algorithms have $\gamma=1$, but in other settings different $\gamma$ may appear.

\section{The mixing lemma}

Here we prove a Lemma that underpins both Thm.~\ref{genThm} and Thm.~\ref{axialThm}, and may also enable further extensions.
\begin{lem}
\label{mainLem}
  Let $V$ be a target unitary, with associated channel $\mathcal{V}(\rho)=V\rho V^\dagger$.  Let $a,b>0$ and $\{ U_1, U_2, \ldots , U_n \}$ be a set of unitaries such that
\begin{enumerate}
	\item for all $j \in \{ 1,\ldots, n \}$	we have $||U_j - V || \leq a$; 
    \item there exist positive numbers $\{ p_j \}$ such that $\sum_{j=1}^n p_j=1$ and $|| (\sum_j p_j U_j) - V || \leq b$.
\end{enumerate}
It follows that $\mathcal{E}=\sum_j p_j \mathcal{U}_j$ satisfies
\begin{equation}
	|| \mathcal{E} - \mathcal{V} ||_{\diamond} \leq a^2+2b	.
\end{equation}
\end{lem}
We will find constructions where $a=O(\epsilon)$ and $b=O(\epsilon^2)$, so that the diamond norm is upper bounded by $O(\epsilon^2)$. 

For now, we prove the above Lemma.  We begin by defining $\delta_j := U_j-V$ so that $|| \delta_j || \leq a$.  We also have
\begin{equation}
		\sum_j p_j \delta_j  = \left( \sum_j p_j U_j \right) - V ,
\end{equation}
with condition (2) of the lemma entailing that $||\sum_j p_j \delta_j|| \leq b$.   The channel $\mathcal{E}$ acts as
\begin{align}
	\mathcal{E}(X) & = \sum_j p_j U_j X U_j^\dagger ,\\ \nonumber	
	& = \sum_j p_j (V + \delta_j) X (V^\dagger + \delta_j^\dagger) .
\end{align}
Since the diamond norm is unitarily invariant, we have $d_\diamond(\mathcal{E},\mathcal{V}) = d_\diamond(\mathcal{V}^\dagger \circ \mathcal{E}, \id)$ where
\begin{align}
	(\mathcal{V}^\dagger \circ \mathcal{E})(X) & = \sum_j p_j V^\dagger U_j X U_j^\dagger V \\ \nonumber	
	& = \sum_j p_j (\id + \tilde{\delta}_j) X (\id  + \tilde{\delta}_j^\dagger) \\ \nonumber 
	& = \sum_j p_j (X + \tilde{\delta}_j X + X \tilde{\delta}_j^\dagger +  \tilde{\delta}_j X \tilde{\delta}_j^\dagger ),
\end{align}
where $\tilde{\delta}_j:=V^\dagger \delta_j$.  Since the operator norm is unitarily invariant, we have $||\sum_j p_j \tilde{\delta}_j|| = ||\sum_j p_j \delta_j|| \leq b$. Compared to the identity channel $\id$, and using $\sum_j p_j =1$, we have
\begin{equation}
	(\mathcal{V}^\dagger \circ \mathcal{E} - \id)(X) = \sum_j p_j ( \tilde{\delta}_j X + X \tilde{\delta}_j^\dagger +  \tilde{\delta}_j X \tilde{\delta}_j^\dagger)	.
\end{equation}
Taking the 1-norm and using the triangle inequality, we have
\begin{align} \nonumber 
	||(\mathcal{V}^\dagger \circ \mathcal{E} - \id)(X)||_1  \leq & ||\sum_j p_j  \tilde{\delta}_j X||_1 + ||\sum_j p_j X \tilde{\delta}_j^\dagger||_1 \\ 
 & + \sum_j p_j  ||\tilde{\delta}_j X \tilde{\delta}_j^\dagger||_1	.
\end{align}
Using the H\"{o}lder inequality and $||X||_1 \leq 1$, we have
\begin{align}
	||(\mathcal{V}^\dagger \circ \mathcal{E} - \id)(X)||_1  & \leq  ||\sum_j p_j  \tilde{\delta}_j || + ||\sum_j p_j \tilde{\delta}_j^\dagger|| \\ \nonumber & + \sum_j p_j  ||\tilde{\delta}_j ||  \cdot || \tilde{\delta}_j^\dagger||	.
\end{align}
Noting the property $|| M || = || M^\dagger ||$ and condition (1) of Lem.~\ref{mainLem}, we conclude that $|| \tilde{\delta}_j^\dagger || = || \tilde{\delta}_j || \leq a$. Therefore, the last sum of terms is upper bounded by $a^2$.  The first two summations are likewise bounded by $b$ by virtue of condition (2).  Therefore,
\begin{align}
	||(\mathcal{V}^\dagger \circ \mathcal{E} - \id)(X)||_1   & \leq  a^2+2b ,
\end{align}
which is true for all $X$.  If we tensor the channels with the identity this does not affect the proof except to burden the notation, and so
\begin{align}
	||((\mathcal{V}^\dagger \circ \mathcal{E}  - \mathcal{I}) \otimes \mathcal{I} )(X)||_1   & \leq  a^2+2b .
\end{align}
Since this is true for all $X$ the diamond norm is also upper bounded by $a^2+2b $. This completes the proof.
\begin{figure*}
	\includegraphics{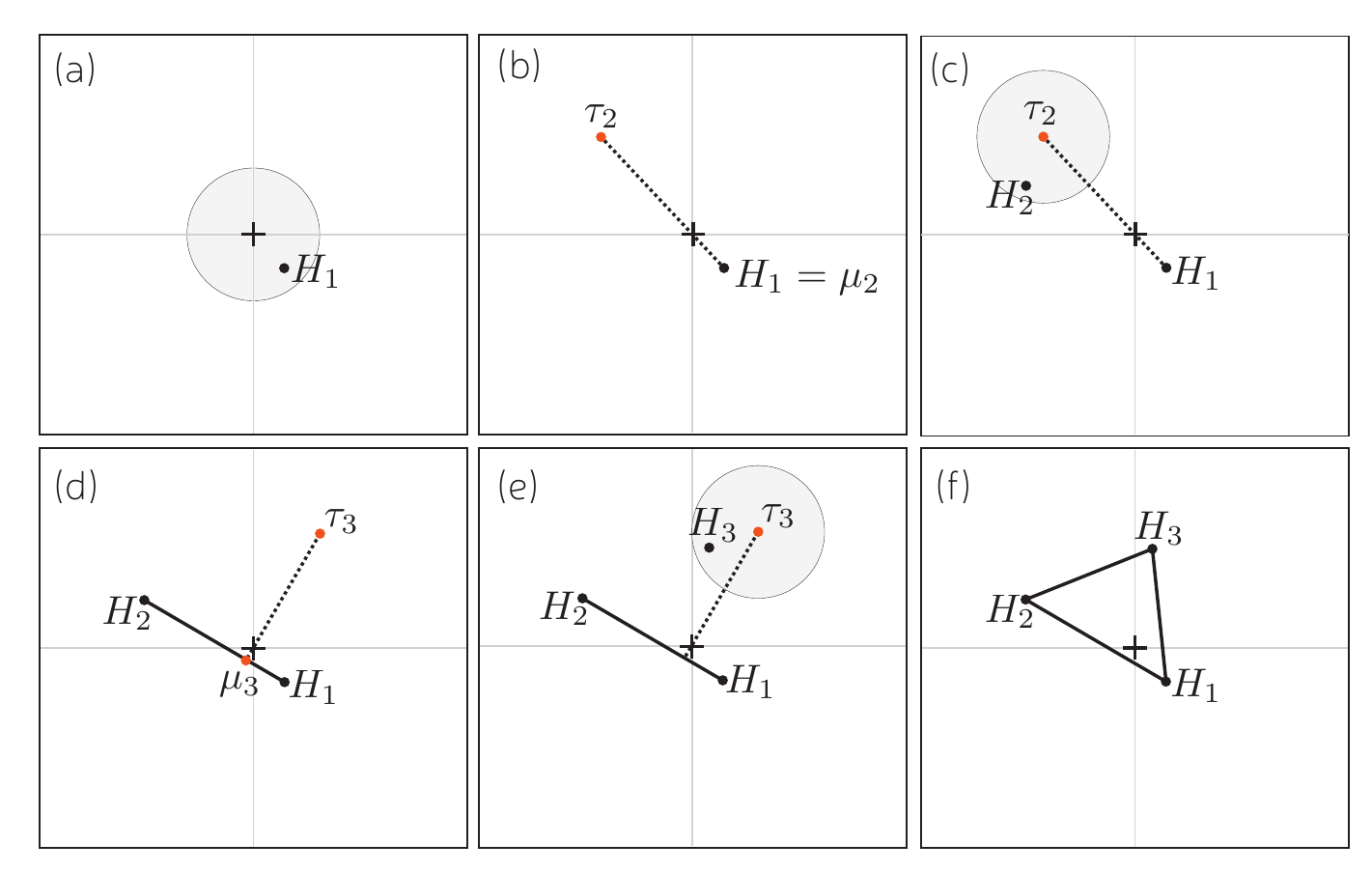}
\caption{The geometric intuition of the convex hull finding algorithm.  The cross marks the origin corresponding to $V$.  (a) We find a $U_1 =  V e^{iH_1}$ so that $H_1$ is near the origin. (b) We extrapolate from $\mu_2=H_1$ through the origin to a point $\tau_2$. (c) We find a $U_2 =  V e^{iH_2}$ close to $Ve^{i\tau_2}$,  so that $H_2$ is near to $\tau_2$.  (d) We form the convex hull of $H_1$ and $H_2$ and find the point $\mu_3$, which is closest to the origin. From here we extrapolate out through the origin to the point $\tau_3$. (e) We find a $U_3 =  V e^{iH_3}$ close to $Ve^{i\tau_3}$, so that $H_3$ is near to $\tau_3$.  (f) We form the convex hull of $H_1, H_2$ and $H_3$ and find the origin lies inside the hull, and so the algorithm terminates. Note that none of the $H_j$ can stray far from the origin.}	
\label{SimplexFig}
\end{figure*}

\section{General rotations}

We show here that Thm.~\ref{genThm} follows from Lem.~\ref{mainLem}.  First, let $\mathcal{G}_\epsilon$ be the subset of $\langle \mathcal{G} \rangle$ such that they can be synthesized with cost not exceeding $f(\epsilon)$.  We have that $\mathcal{G}_\epsilon$ is an $\epsilon$-cover of $\mathcal{L}$.  That is, for all $V \in  \mathcal{L}$ there exists a $U \in \mathcal{G}_\epsilon$ with $|| U - V || \leq \epsilon$.  Since we work with a unitarily invariant norm this can be restated as $|| V^\dagger U - \id || \leq \epsilon$.  We shift to a Hermitian representation and define a $H$ such that $U = Ve^{iH}$. Since $U \sim V$ we can choose $H$ to have small norm, which we verify later.  Our goal is to not just find a single $U$ close to $V$ but a whole set $\{ U_j \}_j$ that allows us to use the following
\begin{lem}
\label{HermLem}
	Let $\{ H_j \}_j$ be a set of bounded Hermitian operators $|| H_j || \leq c$ for all $j$. Assume, the origin lies within the convex hull $0 \in \mathrm{Conv}[ \{ H_j \}_j ]$ with convex decomposition $0 = \sum_j p_j H_j$.  It follows that\begin{enumerate}
	\item $|| e^{i H_j} - \id || \leq c + \frac{c^2}{2}$ for all $j$;
	\item $|| \sum_j p_j  e^{i H_j} - \id || \leq \frac{c^2}{2}$.
\end{enumerate}
When $U_j=V e^{i H_j}$ for some unitary $V$, this can be restated as
\begin{enumerate}
	\item $|| U_j - V || \leq c + \frac{c^2}{2} $ for all $j$;
	\item $|| \sum_j p_j  U_j - V || \leq \frac{c^2}{2}$.
\end{enumerate}
\end{lem}
Clearly, such a set of Hermitian operators would allow us to use Lem.~\ref{mainLem} with constants related by $a=c + \frac{c^2}{2}$  and $b=\frac{c^2}{2}$, yielding an upperbound of $a^2+2b=O(c^2)$.  The lemma is proved by expanding the exponentials into a power series and using standard norm properties, as shown in App.~\ref{HermLemProof}.    

The key point is that we seek a set of Hermitian operators, such that the origin is contained within the convex hull of these points. Next, we present an explicit method for finding such a convex decomposition of Hermitian operators.  We assume access to an oracle performing the relevant gate-synthesis decompositions.  We outline the algorithm for finding a suitable convex set containing the origin.
\\

\noindent
\textbf{Convex hull finding algorithm}
\begin{enumerate}
	\item Call oracle to find $U_1$ such that $||U_1 - V||\leq \epsilon$;
\item Find principle $H_1$ such that $U_1=Ve^{iH_1}$;
\item Set $n=2$ and loop the following
\begin{enumerate}
	\item Find $\mu_n \in \mathrm{Conv}[ \{ H_j \}_{1 \leq j \leq n-1} ] $ with minimum $|| \mu_n ||$;
	\item If $|| \mu_n || = 0$ then EXIT LOOP;
\item Define $W_n = V e^{i \tau_n} $ where $\tau_n:= - r \epsilon \mu_n / ||\mu_n ||$;
\item Call oracle to find $U_n$ such that $||U_n - W_n|| \leq \epsilon$;
\item Find principle $H_n$ such that $U_n=Ve^{iH_n}$ and append to set $\{ H_j \}_{1 \leq j \leq n-1}$;
 \item $n \rightarrow n+1$ and return to start of loop.
\end{enumerate}
\end{enumerate}
The calculation in step 3(a) is a convex optimisation problem and can be solved using standard interior-point methods. The whole algorithm has two free parameters $\epsilon$ and $r$ (see step 3b).  In our analysis we assume $\epsilon \leq 0.01$, and for all practical applications this is easily satisfied.  We take $r=2$ for simplicity, and the exact constants in our bounds and convergence rates depend on this choice.  The algorithm behaves qualitatively the same for different $r$ settings, assuming $\epsilon^{-1} \gg r>1$.  The algorithm has two important properties that we discuss below, leaving technical details until the appendices.  The basic geometric intuition behind the algorithm is illustrated in Fig.~\ref{SimplexFig}.

First, for all $H_j$ found by the algorithm we have
\begin{equation}
	|| H_j || \leq 3 \epsilon + 7 \epsilon^2 ,
\end{equation}
which we show in App.~\ref{HnBound}.  This provides us with the value $c=3 \epsilon + 7 \epsilon^2$ to be substituted into Lem.~\ref{HermLem}, which traced back leads to the diamond norm upper bound
\begin{align}
		d_\diamond( \mathcal{E} , \mathcal{U} ) &\leq \frac{1}{2}(  a^2 + 2b ) = \frac{1}{2} \left[ \left(c + \frac{1}{2}c^2 \right)^2 +  c^2 \right]  \\ \nonumber
& \leq 10 \epsilon^2 ,
\end{align}
where the last line uses $\epsilon<0.01$ to simplify higher order terms.  This gives the upper bound stated in Thm.~\ref{genThm}.

The second important property of the algorithm is that it eventually terminates.  Each $U_n$ is distinct, and in particular its $H_n$ falls outside the convex hull of previous points (see App.~\ref{Converge} for proof).  If we further assume that there are a finite number of distinct points with bounded resource cost, then there are only a finite number of possible $U_n$ for the algorithm to output.  Since each is distinct, the algorithm must terminate in a finite number of steps.  The additional assumption of a finite number of suitable points is very mild, and is satisfied both for the Clifford+$T$ gate set and also any gate set where all gates have non-zero cost.  Furthermore, below we see that the algorithm need not terminate, but that sufficient iterations will work equally well.

A finite number of steps may still be very many, but we have evidence the converge is very fast.  First we note that in a $d$-dimensional space, a simplex of $d+1$ will suffice to enclose a nontrivial volume.  Though the algorithm is not ensured to converge in $d+1$ steps, it may often do so.  Looking at Fig.~\ref{SimplexFig}, the analogous setup in Euclidean geometry hints that it will always find an enclosing simplex in $d+1$ iterations, though it is unclear whether this carries over to the topology induced by the operator norm. We can be more quantitative by considering the quantity $|| \mu_n ||$, which measures the distance from the convex hull.  Recall that the convex hull finding algorithm halts when $|| \mu_n ||=0$.  Further evidence of rapid convergence is that  $|| \mu_n ||$ decreases exponentially fast.   Specifically, we find there exists a $w> 0.62$ such that
\begin{equation}
\label{muMAINbound}
		|| \mu_n || < 6 \epsilon e^{- w n},
\end{equation}
so the convergence toward zero is exponentially fast.  Even exponentially small $|| \mu_n ||$ may be nonzero, but once $|| \mu_n || \ll \epsilon^2$ the preceding proofs can be adapted to account for nonzero $|| \mu_n ||$ with negligible influence on the upper bounds.  All convergence proof details are given in App.~\ref{Converge}.

\section{Axial rotations}

We now consider a setting where the target $V$ is an axial rotation of a single qubit.   The only assumption we make about the generating gate set is that it contains Pauli $Z$ as a free resource.  Given a protocol for axial-synthesis, for all such $V=e^{i \theta Z}$ and any $\epsilon > 0$ there exists at least one $U_1$ such that $||U_1 - V|| \leq \epsilon$ and where $U_1$ has cost not exceeding $f_{\mathrm{ax}}(\epsilon)$ for some $f_{\mathrm{ax}}$.  Recall that $f_{\mathrm{ax}}$ is polylogarithmic in $\epsilon^{-1}$. For instance, the Ross-Selinger algorithm satisfies the worst case bound $f_{\mathrm{ax}}(\epsilon) \leq 4 \log_2 (\frac{1}{\epsilon})$, and $3 \log_2 (\frac{1}{\epsilon})$ on average.  It will prove useful to consider $V^\dagger U_1$ and expand in the Pauli basis
\begin{equation}
	V^\dagger U_1 = \alpha_{\id} \id + i\alpha_{X}X+i\alpha_{Y}Y+i\alpha_{Z}Z.
\end{equation}
We say $U_1$ is an over-rotation if $\alpha_Z \geq 0$ and an under-rotation if $\alpha_Z < 0$.  We require a second unitary $U_2$ such that the pair $\{ U_1 , U_2 \}$ contains one over-rotation and one under-rotation.  We can assume $\alpha_Z \neq 0$ as otherwise the second rotation is not needed.  For the second rotation, we will use the Pauli expansion
\begin{equation}
	V^\dagger U_2 = \beta_{\id} \id + i\beta_{X}X+i\beta_{Y}Y+i\beta_{Z}Z.
\end{equation} 
Gate-synthesis only ensures one unitary such that $||U_1 - V|| \leq \epsilon$, but a suitable $U_2$ can be found only slightly further away. Specifically, there must exist a suitable $U_2$ with cost below $f(\epsilon)$.  To verify this, one first constructs an axial rotation $V'$ with $||V-V'||=\epsilon$ and $|| U_1 - V' ||> \epsilon$.  Specifically, using $V=e^{i \theta Z}$ and $V'=e^{i (\theta + \delta) Z}$ then the two values
\begin{equation}
	\delta = \pm 2 \arcsin( \sqrt{\epsilon} / 2) ,
\end{equation}	
both ensure that $||V-V'||=\epsilon$. Choosing the the sign of $\delta$ to match the sign of $\alpha_{Z}$, it follows that $|| U_1 - V' || > || V- V' ||  =\epsilon$.  Unitary gate synthesis must then provide a $U_2 \neq U_1$ within $\epsilon$ of $V'$, such that $|| U_2 - V || \leq 2 \epsilon$.   Furthermore, within the same cost budget we can synthesize unitaries $U_3 = Z U_1 Z$ and $U_4=Z U_2 Z$, with
\begin{align}
	V^\dagger U_3 & = \alpha_{\id} \id - i\alpha_{X}X - i\alpha_{Y}Y+i\alpha_{Z}Z , \\\nonumber
    V^\dagger U_4 & = \beta_{\id} \id - i\beta_{X}X-i\beta_{Y}Y+i\beta_{Z}Z.
\end{align}
Considering the set $\{  U_1, U_2, U_3, U_4 \}$ it follows immediately that they satisfy condition (1) of Lem.~\ref{mainLem} with $a =  2\epsilon$.  Next, we assign them weights $\{ p_j \} = \{ \frac{1-q}{2}, \frac{q}{2}, \frac{1-q}{2}, \frac{q}{2}  \}$ where $0 \leq q\leq 1$ will be fixed later. The linear combination is
\begin{align}
	\sum_j p_j V^\dagger U_j = & ((1-q) \alpha_\id + q\beta_\id ) \id \\ \nonumber 
 & + i((1-q) \alpha_Z + q\beta_Z ) Z.	
\end{align}
Subtracting the identity and taking the operator-norm squared,
\begin{align}
	||\sum_j p_j V^\dagger U_j - \id || ^2 = & ((1-q) \alpha_\id +q \beta_\id -1 )^2\\ \nonumber 
 & + ((1-q) \alpha_Z +q \beta_Z )^2	.
\end{align}
We now fix $q$ to eliminate the second term.  Considering the variables $\{ \alpha_Z$, $\beta_Z \},$ one is positive (an over-rotation) and the other negative (an under-rotation), so zero sits within the convex hull of these variables and suitable $q$ can be found. Specifically
\begin{equation}
	q = \frac{\alpha_Z}{\alpha_Z - \beta_Z} ,
\end{equation}
satisfies $0 \leq q \leq 1$. With the second term cancelled and taking square roots we have
\begin{align}
	||\sum_j p_j V^\dagger U_j - \id || & =  \left| q (\beta_\id-\alpha_\id) +(1-\alpha_\id) \right|.
\end{align}
By the triangle inequality and $|q| \leq 1$, we have
\begin{align}
	||\sum_j p_j V^\dagger U_j - \id || & \leq   |  \beta_\id-\alpha_\id |+|\alpha_\id-1|.
\end{align}
Inserting $1-1=0$, so that $ \beta_\id-\alpha_\id =  (\beta_\id-1)+(1-\alpha_\id)$, and again using the triangle inequality, we arrive at
\begin{align}
\label{eq::midway}
	||\sum_j p_j V^\dagger U_j - \id || & \leq   | \beta_\id - 1|+ 2|\alpha_\id-1|.
\end{align}
From $||V^\dagger U_1 - \id || \leq \epsilon$ we can infer that
\begin{equation}
 || (\alpha_\id -1)\id + i\alpha_{X}X+i\alpha_{Y}Y+i\alpha_{Z}Z ||^2 \leq \epsilon^2.
\end{equation}
Evaluating the left hand side, we obtain
\begin{equation}
(\alpha_\id -1)^2+\alpha_{X}^2+\alpha_{Y}^2+\alpha_{Z}^2 \leq \epsilon^2.
\end{equation}
Unitarity of $V^\dagger U_1$ entails that $\alpha_\id^2+\alpha_{X}^2+\alpha_{Y}^2+\alpha_{Z}^2=1$ and  after some simplification, we find
\begin{align} \nonumber
 (\alpha_\id -1)^2+\alpha_{X}^2+\alpha_{Y}^2+\alpha_{Z}^2 & = (\alpha_\id -1)^2+(1-\alpha_\id^2), \\ 
 & = 2(1-\alpha_\id) \leq \epsilon^2. 
\end{align}
From which we infer $|1-\alpha_\id | \leq \epsilon^2 /2$.  Similarly, from $||V^\dagger U_2 - \id || \leq 2\epsilon$ we can infer $|1-\beta_\id | \leq 2 \epsilon^2$.  Substituting into Eq.~\eqref{eq::midway}, we have 
\begin{equation}
		||\sum_j p_j V^\dagger U_j - \id || \leq 3 \epsilon^2 ,
\end{equation}
Therefore, we have demonstrated both the necessary conditions of Lem.~\ref{mainLem} with $a=2\epsilon$ and $b=3 \epsilon^2$.  Applying the Lemma, our channel satisfies
\begin{equation}
	d_\diamond( \mathcal{E} , \mathcal{V} ) \leq \frac{1}{2}(a^2+2b) \leq 5 \epsilon^5.
\end{equation}
A smaller factor than 5 is likely to be provable.

\section{Conclusions}

We have seen that worst-case resource costs of fault-tolerant quantum computing can be reduced by switching to a randomised approach to gate-synthesis.  It may seem counterintuitive that a randomisation process can be advantageous.  However, convexity of the diamond distance naturally entails that mixing over channels of similar noise levels can only reduce the noise.  

We presented a convex hull finding algorithm for finding the suitable mixing ratios. While this algorithm is exponentially fast, it is plausible that a constant time algorithm exists.  We suspect that a variant of Delaunay triangulation could be used to quickly identify a suitable simplex.  However, our literature search on Delaunay triangulation has only found results on Euclidean space and we have yet to ascertain if such tools carry over to the operator norm topology.

This work has only considered mixing over unitary channels, which prompts the question whether more general quantum channels might be useful.  Probabilistic quantum circuits with fallback~\cite{bocharov15} is an approach to gate-synthesis that is not entirely unitary, though it makes use of an ancillary qubit and works very differently to the approach presented here.  As remarked earlier, mixing can be useful in preparation of different magic states~\cite{campbell16}.  We ponder whether all these approaches can be understood within a single framework of quantum channel synthesis.  

Acknowledgements.- We thank Yuan Su, Vadym Kliuchnikov, Steve Flammia, Jens Eisert, Mark Howard, Luke Heyfron, Neil J. Ross, Mark Pearce and Scott Vinay for related discussions. Several of these discussions were facilitated by the Centro de Ciencias de Benasque. This work was supported by the EPSRC (EP/M024261/1).

\appendix

\section{Convex hull proof}
\label{HermLemProof}

This section will prove Lem.~\ref{HermLem}.   We start by showing another general result that we use in several places.   Let $M$ be a Hermitian operator with eigenvalues $\lambda_k$, so that by definition $|\lambda_k| \leq ||M||$ for all $k$.  We consider the operator  
 \begin{align}
     e^{i M} - (\id+iM)  & =  \sum_{n=2}^{\infty} \frac{1}{n!}(i M)^n  .
 \end{align}
This can be diagonalised in the eigenbasis of $M$ and has eigenvalues $f_M(\lambda_k):=e^{i \lambda_k}-1-i\lambda_k$.   Therefore, we have  
 \begin{align}
 ||  e^{i M} - (\id+iM) || =  \mathrm{max}_{k} |f_{M} (\lambda_k) |.
\end{align}
On the interval $|x| \leq \pi$, one can verify that $|e^{ix}-1-ix| \leq \frac{1}{2} x^2$, and so provided $||M|| \leq \pi$ we have 
 \begin{align}
 \label{Mquadratic}
	||  e^{i M} - (\id+iM) || \leq \frac{1}{2} ||M||^2.
\end{align}
Now turning specifically to Lem.~\ref{HermLem}, we have 
\begin{align}
		|| e^{i H_j} - \id || & = ||\sum_{n=1}^{\infty} \frac{1}{n!}(iH_j)^n || \\
& \leq ||H_j|| + ||\sum_{n=2}^{\infty} \frac{1}{n!}(i H_j)^n || . 
\end{align}
Since we always choose the principle $H_j$, we have $||H_j||\leq \pi$ and we can use Eq.~\ref{Mquadratic} to find
\begin{align}
|| e^{i H_j} - \id || & \leq ||H_j|| + \frac{1}{2}||H_j ||^2  \leq c + \frac{1}{2}c^2. 
\end{align}
Recall that in  Lem.~\ref{HermLem} we defined $c$ so that $||H_j|| \leq c$ for all $H_j$, which explains the second inequality.   Therefore, $|| e^{i H_j}  || \leq c + \frac{c^2}{2}$.  This shows property (1) of Lem.~\ref{HermLem}.  Next we consider the convex sum of unitaries, 
\begin{align}
	\sum_j p_j  e^{iH_j} & =   \id + (\sum_j  ip_j H_j) + \sum_j p_j \sum_{n=2}^\infty \frac{(iH_j)^n}{n!},
\end{align}
which is split into zeroth, first and higher order terms.  By assumption the linear terms vanish.  Therefore,
\begin{align}
	|| \sum_j p_j  e^{iH_j} - \id || = &  ||\sum_j p_j \sum_{n=2}^\infty \frac{(iH_j)^n}{n!}||, \\ \nonumber
 & \leq \sum_j p_j || \sum_{n=2}^\infty \frac{(iH_j)^n}{n!}|| \\ \nonumber
 & \leq \sum_j p_j  \frac{c^2}{2} = \frac{c^2}{2} .
\end{align}
Going from second to third line, we have again used Eq.~\ref{Mquadratic}.  This proves Lem.~\ref{HermLem}.

\section{Bounding $||H_n ||$.}
\label{HnBound}

We wish to upper bound $||H_n||$ in terms of $\epsilon$, the precision to which gate synthesis is assessed.  The operator $H_n$ is chosen so that $e^{iH_n}$ provides a certain unitary, $U_n$, and the eigenvalues are chosen within the interval $[ -\pi ,  \pi)$.  Furthermore, on this interval one has that all eigenvalues $\theta$ satisfy $|\theta| \leq  | e^{i \theta} - 1 |+ \frac{1}{2}| e^{i \theta} - 1 |^2$.  It follows that 
\begin{equation}
\label{HnUpperRough}
	||H_n|| \leq || e^{i H_n} - \id || + \frac{1}{2}|| e^{i H_n} - \id ||^2 .
\end{equation}
Next, we note that for each $n>1$ we have
\begin{align}
	|| e^{iH_n} - \id || & = || U_n - V || \\ \nonumber
& \leq || U_n - W_n || + || W_n - V || \\ \nonumber
 & \leq \epsilon + || e^{i \tau_n} - \id || \\ \nonumber
 & \leq \epsilon + || \tau_n|| + \frac{|| \tau_n||^2}{2} \\ \nonumber
 & \leq 3\epsilon + 2 \epsilon^2. 
\end{align}
The $n=1$ case is similar but without the  $|| W_n - V ||$ contribution.  Combining this with Eq.~\eqref{HnUpperRough} we have
\begin{equation}
		||H_n || \leq \left(  3\epsilon + 2 \epsilon^2 \right) + \frac{1}{2}\left(  3\epsilon + 2 \epsilon^2 \right)^2.
\end{equation}
Assuming $\epsilon < 0.01$ this can be simplified to
\begin{equation}
		||H_n || \leq 3 \epsilon + 7 \epsilon^2,
\end{equation}
as reported in the main text.  This gives the value of $c$ for Lem.~\ref{HermLem}.

\section{Convergence proof}
\label{Converge}

Next we show that each $U_n$ is new by showing the strictly monotonic decrease of $|| \mu_n ||$.  Furthermore, we show exponential decrease of $|| \mu_n ||$ with $n$.  We begin by translating the closeness of $U_n$ to $W_n$ into the space of Hermitian operators.  We define
\begin{equation}
	\Delta_n := H_n - \tau_n,	
\end{equation}
and later will find an upper bound on $|| \Delta_n ||$.  First we use these operators to construct a point in the new convex hull.  Mixing $H_n$ and $\mu_n$ gives a point in the convex hull, which must have norm no larger than $|| \mu_{n+1} ||$, so that
\begin{align}
	||\mu_{n+1}||	& \leq || \lambda H_n + (1-\lambda) \mu_n || \\ \nonumber
& = || \mu_n \left[ 1 - \lambda \left( 1 + 2 \frac{\epsilon}{|| \mu_n ||} \right) \right]  + \lambda \Delta_n ||.
\end{align}
If we consider when
\begin{equation}
	\lambda= 	\left( 1 + 2 \frac{\epsilon}{|| \mu_n ||} \right)^{-1} = \frac{|| \mu_n ||}{|| \mu_n || + 2 \epsilon },
\end{equation}
then it is easy to see $0 < \lambda < 1$ and that the square bracket vanishes so that
\begin{align}
\label{muITERATE}
	||\mu_{n+1}|| & \leq  \lambda || \Delta_n || =  \frac{|| \mu_n ||}{|| \mu_n || + 2 \epsilon }|| \Delta_n ||.
\end{align}
This iteration begins with $\mu_2=H_1$.  Further progress requires an upper bound on $|| \Delta_n ||$, which we now take a lengthy detour to find.

Adding several terms of the form $0=(x-x)$ to $\Delta_n$, we have
\begin{align}
	\Delta_n  = & (-i\id + H_n + ie^{iH_n})+(iV^\dagger W_n-ie^{iH_n}) \\ 
 	&+\left( -iV^\dagger W_n + i\id - \tau_n \right)	 . \nonumber
\end{align}
Taking the norm and applying triangle inequality, we get
\begin{align}
	|| \Delta_n ||  \leq & ||\id + iH_n -e^{iH_n}||+||V^\dagger W_n -e^{iH_n}|| \\ 
 	&+||\id + i\tau_n - V^\dagger W_n   ||	 \nonumber \\ \nonumber
= & ||\id + iH_n -e^{iH_n}||+||W_n - U_n|| \\ 
 	&+||\id +i\tau_n - e^{\tau_n}   ||	.
\end{align}
For the middle term we know $||W_n - U_n||\leq \epsilon$, and for the first and last terms we  again use Eq.~\ref{Mquadratic}, so that  
\begin{align}
	|| \Delta_n ||  &\leq \frac{1}{2}||H_n||^2+ \epsilon + \frac{1}{2}|| \tau_n||^2 \\ \nonumber
 & \leq \frac{1}{2} (3 \epsilon + 7 \epsilon^2)^2 + \epsilon + 
 \frac{1}{2} (2 \epsilon)^2.
\end{align}
We can again use $\epsilon \leq 0.01$ to bound higher order terms to obtain
\begin{align}
	|| \Delta_n ||  &\leq \epsilon + 7 \epsilon ^2 .
\end{align}
Plugging this in Eq.~\eqref{muITERATE}, we have
\begin{align}
	||\mu_{n+1}||	&  \leq  || \mu_n || \frac{ \epsilon + 7 \epsilon^2}{|| \mu_n || + 2 \epsilon }   \\ \nonumber
& <  || \mu_n || \frac{1}{2}\left( 1 + 7 \epsilon \right) ,
\end{align}
where we have used that $0<|| \mu_n ||$.  Iterating this argument $n$ times we find exponential behaviour
\begin{align}
	||\mu_{n+1}||	& <  || \mu_2 || e^{- w  (n-1) } ,
\end{align}
where $w=\ln(2) - \ln(1+7 \epsilon)$.  Using our earlier assumption that $0<\epsilon < 0.01$  guarantees that $0.69315>w>0.62548$.  In most instances convergence will be much faster than ensured by this proof, often jumping to $||\mu_{n+1}||=0$ within only a few iterations.  Last we note that $\mu_1 = H_1$ and that $|| H_1 || \leq  || V- U_1 ||+\frac{1}{2}|| V- U_1 ||^2 \leq \epsilon + \frac{1}{2}\epsilon^2$, which gives 
\begin{align}
	||\mu_{n}|| & < \epsilon \left( 1 + \frac{1}{2}\epsilon \right) e^{- w (n-2) }	\\ \nonumber
	& =   \epsilon  \left[ \left( 1 + \frac{1}{2}\epsilon \right)e^{2w}  \right] e^{- w n}	 .
\end{align}
Since $e^{-w} > 1/2$, we have $e^{2w} < 4$. Combined with $\epsilon < 0.01$ we know the square bracket cannot exceed 6, which leads to Eq.\eqref{muMAINbound}.


%

\end{document}